\documentstyle[12pt]{article}

\begin{document}

\renewcommand{\thesection}{\arabic{section}.}
\renewcommand{\theequation}{\thesection \arabic{equation}}
\newcommand{\scs}{\setcounter{equation}{0} \setcounter{section}}

\def\req#1{(\ref{#1})}
\newcommand{\del} {\partial}
\newcommand{\ha} {{1\over 2}}
\newcommand{\ov} {\over}
\newcommand{\tvp} {{\tilde \vp}}
\newcommand{\vp} {{\varphi}}
\newcommand{\p} {\phi}
\newcommand{\fourth} {{1\over 4}}
\newcommand{\be}{\begin{equation}}
\newcommand{\ee}{\end{equation}}
\newcommand{\ba}{\begin{eqnarray}}
\newcommand{\ea}{\end{eqnarray}}

\newcommand{\la}{\label}

\newcommand{\lb}{\large\bf}

\newcommand{\bb} {\bibitem}
\newcommand{\np} {{\it Nucl. Phys. }}
\newcommand{\pl} {{\it Phys. Lett. }}
\newcommand{\pr} {{\it Phys. Rev. }}
\newcommand{\mpl} {{\it Mod. Phys. Lett. }}

\begin{titlepage}
\renewcommand{\thefootnote}{\fnsymbol{footnote}}

\hfill BUTP-94/7

\hfill Imperial/TP/93-94/29

\hfill hep-th/9404180

\vspace{.4truein}
\begin{center}
 {\LARGE On string cosmology and the RG flow}

 {\LARGE in $2d$ field theory }
\end{center}

\vspace{.4truein}
\begin{center}
C. Schmidhuber\footnote{e-mail: christof@butp.unibe.ch}  \vskip1mm
 {\it Institute for Theoretical Physics}

 {\it University of Bern, CH-3012 Bern, Switzerland}
\vskip5mm {\it and}\vskip5mm
 A.A.  Tseytlin\footnote{e-mail: tseytlin@ic.ac.uk.
\  On leave  from Lebedev  Physics Institute, Moscow, Russia}

 {\it  Theoretical Physics Group, Blackett Laboratory}

 {\it  Imperial College,  London SW7 2BZ, U.K. }

\end{center}

\vspace{.3truein}
\begin{abstract}

Time--dependent solutions of bosonic string theory resemble
renormalisation group trajectories in the space of $2d$ field theories: they
often interpolate between repulsive and attractive static solutions.
It is shown that the attractive static solutions are those
whose spatial sections are minima  of $\vert \bar c-25 \vert$,
where $\bar c$ is the `$c$-function'.
The size of the domain of attraction of such a solution may be a measure
of the probability of the corresponding string vacuum.
Our discussion  has also  an implication for the  RG
flow in theories coupled to dynamical $2d$ gravity:
the  flow from models with $c>25$ to models with $c<25$ is forbidden.

\end{abstract}
\vfill\hskip1cm{4/94}\vskip1cm
\renewcommand{\thefootnote}{\arabic{footnote}}
\setcounter{footnote}{0}
\end{titlepage}

{}\section*{\lb 1. Introduction}\scs{1}

If the universe is described by string theory, the central charges of the flat
spatial coordinates and of the compactification space
must add up to $c_{tot} = 25$ ($\hat c_{tot}=9$ in superstring theory)
at the present stage of evolution. Otherwise we would observe a time--dependent
dilaton background \cite{Myr,Anto}, which -- from the world--sheet point of
view -- is needed
to balance the central charge and which would manifest
itself in  time dependence of  coupling constants.

However, one may speculate that $c_{tot}$ has been larger or smaller than 25 in
the early universe.\footnote{Such a suggestion was made in \cite{Anto}.}
In the present paper we derive a result that
supports this proposal. Namely,
the space of general time-dependent solutions of classical bosonic string
theory contains attractors:
they are static solutions with `spatial' CFT parts  which  are minima of $\vert
\bar c-25\vert$, where $\bar c$ is a version  of Zamolodchikov's
$c$--function \cite{zam} equal to the central charge  $\bar c = c_{tot}$  at
its extrema.
One reason why this is  potentially interesting  is that it seems to suggest a
natural way to assign different probabilities
to different compactifications -- according to the sizes of the corresponding
domains of attraction.  Though there is no energy-type criterion for comparing
different compact  internal spaces with the same central charge, by
considering  their possible
time evolution in the early universe
one can thus hope to be able to  determine which of them are `most probable'
ones.

To simplify the problem, we  shall assume that the evolution of the three
`large' spatial dimensions is decoupled and can be ignored.
What we call attractors are thus static attractors for the compact internal
space.\footnote{The evolution of the three ``large dimensions'' could in
principle be included in the discussion.
There may also be non-static attractor solutions; they are not considered
here.}
In contrast to most previous discussions of the cosmological evolution of the
internal space
(see e.g. \cite{Mull}), in which only the moduli (e.g.,  radii of
toroidal compactification) change in time,
we are interested in solutions  in  which the central charge of the internal
conformal field theory (CFT) changes.
We shall discuss bosonic string theory and ignore the tachyon as usual since it
is absent in the superstring generalisation. We shall consider only the leading
 order in the string coupling expansion, i.e., classical string theory,
ignoring  possible (non-)perturbative corrections like a dilaton potential.

To derive the above result we shall follow an analogy with the standard
renormalisation group flow. As discussed in sections 2 and 3,
a generic time-dependent classical solution of the  string  field equations has
much in common with a generic
RG trajectory in the space of $2d$ field theories.
Classical solutions of string theory  can  also be viewed as trajectories in
the $2d$ theory space.
For example, a solution of $N+1$ dimensional bosonic string low--energy
effective equations is given by the target space fields
($\mu,\nu=0,1,...,N+1$)
$$G_{\mu\nu}(\vec x,t), \  B_{\mu\nu}(\vec x,t), \   \phi(\vec x,t)$$
where $x^i \ (i=1,...,N)$ are the  spatial coordinates, $t$ is time,
$G_{\mu\nu}$ is the target space metric,
$B_{\mu\nu}$ is the antisymmetric tensor field, and $\phi$ is the
dilaton.\footnote{For simplicity, we shall ignore all other
possible couplings. Our  argument can be repeated in more general terms, using
perturbative string field theory \cite{sen} or
Wilson's RG approach \cite{bank}.}
At least locally, diffeomorphism symmetry and the gauge symmetry associated
with the antisymmetric tensor field can be used to set
\begin{eqnarray}
  B_{0i}=0,\ G_{00}=\pm1,\ G_{0i}=0.
\label{B1}\end{eqnarray}
Then an $N+1$ dimensional string solution can be  represented as a trajectory
$ \vec\lambda (t)$  (with time being the parameter along the trajectory) in the
space of $N$--dimensional fields
$$ \vec\lambda=\{G_{ij}(\vec x),B_{ij}(\vec x),\phi(\vec x)\}$$
 which is the space of coupling constants of $2d$ bosonic sigma models with a
compact $N$--dimensional euclidean target space  and a curved background world
sheet metric \cite{lov,frts,fri}
\begin{equation}
 I= {1\over{4\pi\alpha'}}\int d^2\xi [ {\sqrt g}
  G_{ij}(\vec x)\partial_\alpha x^i\partial^\alpha x^j  +
 i\epsilon^{\alpha\beta} B_{ij}(\vec x)\partial_\alpha x^i\partial_\beta x^j
  +\alpha' {\sqrt g}R^{(2)}\phi(\vec x)] \ .
\label{A0}\end{equation}
The orbits  $\vec\lambda (t)$  will be compared with the  standard RG
trajectories  in  the sigma model (\ref{A0}).
The  RG trajectories interpolate between   stable and unstable fixed points in
theory space.  A generic RG trajectory is attracted
to some stable fixed point. As will be explained, similar statements hold for
the string solutions $\vec\lambda(t)$ because of the
`friction' provided by the time-dependent  dilaton (see also
\cite{suss,tsey,mukh}\footnote{For a different discussion of a `friction' in
string theory, an interpolation between fixed points, and the RG flow, see
\cite{ellis}.}).

The similarity between the RG flow (satisfying the standard first-order RG
equations) and time-dependent string solutions
(satisfying  second-order equations)  that we shall exploit here has another
interesting interpretation:
a  string  solution $\vec\lambda(t)$  represents an RG flow  in the presence of
{\it $2d$ gravity} \cite{kpz}.\footnote{For  discussions of a relation between
the RG flow with dynamical $2d$ gravity and string theory see, e.g.,
\cite{das,bankslykk,tseyt,suss,sen,pkov} and also
\cite{sch,pkk,amb,tanii,ellis}.}
We shall find  an interesting qualitative effect of coupling a $2d$ field
theory to $2d$ gravity:
it makes it impossible to flow from fixed points
with $c>25$ to fixed points with $c<25$.\footnote{ It makes sense to discuss
CFT's with $c>25$ coupled to gravity in the
approach of  \cite{d,dk}, e.g.,  as  bosonic sectors of superconformal field
theories coupled to supergravity.  One should  not worry
about the `wrong' sign of the kinetic term of the Liouville factor $t$ for
$\bar c>25$: the associated ghosts  should decouple
according to the no--ghost theorem for critical string theory ($\bar
c+c_t=26$).
}

This paper is organised as follows.  In section 2 the string equations of
motion
 are represented as equations for $\vec\lambda(t)$. They take a simple form if
expressed in terms of the  $\beta$-functions
of the spatial $2d$ field theory. Although obtained  to lowest order in
$\alpha'$, these equations should be exact in $\alpha'$
at  the vicinity of  static  solutions. In section 3 the equations are compared
with the RG flow equations and the result stated
above is derived. In section 4  we  consider   implications for the  RG  flow
in the presence of   $2d$        gravity.
As an illustration of the preceding discussion, the example of  the  group
space  sigma model with a Wess--Zumino term is  considered in section 5.
The corresponding solutions  describe an interpolation between flat and
compactified space.

Section 6 contains some concluding remarks.

{}\section*{\lb 2. Equations of motion for the couplings $\vec \lambda (t)$
}\scs{2}

If a string  solution $\vec\lambda(t)$ asymptotically approaches a static
solution (`fixed point'), $\vec\lambda(t)$  obeys a simple
equation of motion at the vicinity of this fixed point, which will be derived
below.

\subsection*{\normalsize\bf 2.1. String effective field equations and
$\beta$-functions}

Consider the world--sheet theory (\ref{A0}).
The  corresponding leading-order
$\beta$-functions  (Weyl anomaly coefficients) for $G_{ij}, B_{ij}$ and $\phi$
can be represented in the form \cite{fri,frts}
\begin{eqnarray}
  \beta^{G(N)}_{ij}&=& \alpha'(
R_{ij}+2\nabla_{i}\nabla_{j}\phi-{1\over4}H_{imn} H_{j}^{mn}) \   ,\label{C1}\\
  \beta^{B(N)}_{{ij}}&=&\alpha'(-{1\over2}\nabla_m H^m_{ij}
+H^m_{ij}\del_m\phi)\    ,\label{C2}\\
  \tilde \beta^{\phi (N)}&=&\beta^{\phi (N)}- {1\over 4} G^{ij}   \beta^{G
(N)}_{ij}=  {1\over6}[C^{(N)}(\vec x)-26]\   .\label{C3}
\end{eqnarray}
Here $H_{ijk}=3\nabla_{[i}B_{jk]} $ and the function
\begin{eqnarray}
  C^{(N)}(\vec x)&=&N-{3\over
2}\alpha'[R-{1\over12}H^2-4(\nabla\phi)^2+4\Box\phi] \label{C4}
\end{eqnarray}
becomes  $\vec x$--independent and equal to the central charge when  the sigma
model  represents   a  conformal theory \cite{CP}.
Eqs.(\ref{C1})--(\ref{C3}) can be derived from the  $N$--dimensional target
space action  \cite{frts,fri} (for a review see \cite{SM})
\begin{eqnarray}
   S^{(N)}=\int d^Nx{\sqrt G}\ e^{-2\phi}[ C^{(N)}(\vec x)-26] \  .
\label{C5}\end{eqnarray}
The string effective equations for $\vec\lambda(t)$ are the requirements that
the $\beta$-functions of the sigma model with
$N+1$ dimensional target space (and
the additional couplings in  (\ref{B1}))
vanish.  Writing (\ref{C1})--(\ref{C3}) in $N+1$ dimensions and making the
$N+1$ split
we find the following expressions in terms of the $\beta$-functions of the
sigma model with $N$--dimensional
target space (in what follows we set $\alpha'=2$):
\begin{eqnarray}
  0 =\beta_{00}^{G (N+1)}&=& 2\ddot \varphi - {1\over2} G^{ik} G^{jl} (\dot
G_{ij} \dot G_{kl } +  \dot B_{ij} \dot B_{kl })\ ,\label{D1}\\
  0 =\beta_{ij}^{G (N+1)}&=&\beta_{ij}^{G (N)} -G^{00} [ \ddot G_{ij} -  \dot
\varphi \dot G_{ij} -
   G^{mn} (\dot G_{im}\dot G_{jn} - \dot B_{im}\dot B_{jn})] \ , \label{D2}\\
  0 =\beta_{ij}^{B (N+1)}&=&\beta_{ij}^{B (N)}
  -G^{00}( \ddot B_{ij} -  \dot \varphi \dot B_{ij}  - 2 G^{kl} \dot G_{k[i}
\dot B_{j]l}  ) \ , \label{D3}\\
  0 =C^{(N+1)}-26         &=&C^{(N)}-25 -3G^{00}( \ddot \varphi - \dot
\varphi^2), \label{D4}
\end{eqnarray}
where we introduced the  shifted dilaton
\[ \varphi = 2\phi - \log {\sqrt G}. \]
The dilaton  equation \req{D4} can also be written in terms of the
$\beta$-function
$\beta^{\phi (N)}$ in \req{C3} to keep the analogy with \req{D2},\req{D3}.
The above    equations follow from the action\footnote{In the simplest case of
homogeneous $x^i$-independent fields the  resulting system of equations is the
same as the $O(N,N)$ duality invariant system in \cite{vene}.}
\[
   S^{(N+1)}=\int dtd^Nx\sqrt{ |G^{00}|} \ e^{-\vp}\ \{ C^{(N)}(\vec x)-25 + {3
} G^{00}
[\dot \varphi^2 -  \fourth G^{ik} G^{jl} (\dot G_{ij} \dot G_{kl } +  \dot
B_{ij} \dot B_{kl }) ] \}    .
\label{CC5}\]
A combination of  (\ref{D1}) and  (\ref{D4}) is a  constraint. The  additional
two constraints
$$ 0=\beta_{0i}^{G (N+1)}=   \nabla_k(G^{kl} \dot G_{li})  -  {1\over 2} \dot
B_{kl} H_i^{\ kl}  + 2 \del_i \dot \varphi -
G^{kl} \dot G_{li} \del_k \varphi  \ ,   $$
$$ 0=\beta_{0i}^{B (N+1)}=- G_{ik}\del_j(G^{kl}G^{jn}\dot B_{nl})   + 2   \dot
B_{ij}\nabla^j\varphi \ ,  $$
 further restrict the initial values of the fields and their time derivatives.
  We shall assume that these constraints can be ignored in the context of our
discussion.\footnote{This is  clear if   the
 constraints on fields and their first time derivatives can be solved for
arbitrary $G_{ij}$,$B_{ij}$,
 $\phi$, which is likely to be the case in perturbation theory in $t$  near the
fixed points that we shall mostly consider. }

It is useful for the following to split the shifted dilaton
$\varphi(\vec x,t)$ into an $\vec x$--dependent part $\tilde\varphi(\vec x,t)$
and an $\vec x$--independent part $\varphi_0(t)$ as follows:
\be
\label{d1}
 \varphi_0(t)\equiv -  \log [\int d^Nx\ e^{-\varphi(\vec x,t)}]\ ,\ \ \  \
   \tilde\varphi(\vec x,t)\equiv\varphi(\vec x,t)-\varphi_0(t)\ .
\ee
 $\varphi_0$ is  thus  minus the logarithm of the  proper  space volume (at
fixed $t$)
\be
\label{d2}
 V^{(N)}\equiv  \int d^Nx{\sqrt G}\ e^{-2\phi} =  e^{- \varphi_0}  \ .
\ee
Let us   define the  space average of a function $f(\vec x)$ by
\begin{eqnarray}
  <f(\vec x)> \equiv {{\int d^Nx\ f(\vec x)\ e^{-\varphi(\vec x)}}\over
  {\int d^Nx\ e^{-\varphi(\vec x)}}} \ .
\label{E1}\end{eqnarray}
An important role is played by  the function
\begin{eqnarray}
Q(t)\equiv-\dot\varphi_0(t) = - <\dot\varphi(\vec x,t)> \ .
\label{EE1}\end{eqnarray}
As follows from (\ref{D1}),  at  fixed points  $\dot G_{ij}=\dot B_{ij}=0 $,
$\ddot \varphi =\ddot  \phi=0 $
so that  $Q$ is constant and  $\varphi_0 (t) = -{1\over2}Q t + const $, but in
general $Q$ will depend on $t$.
Integrating (\ref{D4}) weighted by $e^{-\varphi}$  yields
\begin{eqnarray}
  \dot Q+Q^2=- {1\over3}G^{00}  (\bar c-25),
\label{A2}\end{eqnarray}
where (\ref{E1}) has been used and we have defined  the function
\begin{eqnarray}
\bar c (t) = \bar c(G,B,\varphi) = <C^{(N)}(\vec x)> ={ S^{(N)}\over
V^{(N)}}+26\ .
\label{AA2}\end{eqnarray}
$ S^{(N)}$ is the action in  (\ref{C5}).
 $\bar c$  (or, more precisely,  $ S^{(N)}$)
can be considered a generalisation of the `$c$--function' \cite{zam} to the
case of  sigma models with dilaton coupling \cite{tsy,os}.\footnote{Note that
if  $\vp$ is used as
the independent dilaton coupling then the equations for $G_{ij}$ and $B_{ij}$
that follow from the variation of  $\bar c$   and  $ S^{(N)}$ are equivalent.}
Eq. (\ref{D1}) implies that
\begin{eqnarray}
  \dot Q &=& < -\ddot\varphi + \dot{\varphi}^2 - Q^2> = < (\dot{\varphi} -
<\dot{\varphi}>)^2 -\ddot\varphi >\\
  &=&< \dot{\tilde\varphi}^2-{1\over4} G^{ik} G^{jl} (\dot G_{ij} \dot G_{kl }
+  \dot B_{ij} \dot B_{kl }) > .
\label{A3}\end{eqnarray}

\subsection*{\normalsize\bf 2.2.  Equations  at  the vicinity of fixed points}

In the following we shall study  the behavior of solutions which
asymptotically approach fixed points.
By a fixed point of the above equations we mean a solution in which all
time derivatives are zero, except for $\dot\varphi_0 $   or (what is the same
at a fixed point)
$\dot\phi_0$, where $\phi_0$ is the spatially constant mode of the dilaton.
We thus   allow for a possible   $\vec x$--independent dilaton background
that grows linearly in time (the sigma model with $N$--dimensional
target space  obviously does not `feel' an $\vec x$--independent dilaton
background).

First let us consider the case when   $ c\neq25$.
Within an appropriate neighbourhood of the fixed point, both $\bar c(G,B,\phi)$
and $C^{(N)}(x)$ can be approximated by a constant $c$.
As follows from (\ref{D1}),(\ref{A3}), to lowest order
$\ddot\varphi=0, \ddot{\tilde\varphi}=0$, $\dot Q=0$.  Eq.(\ref{A2}) then
reduces to
\[  Q^2 = -{ 1\over3}G^{00}{(c-25)}.\]
In order to have a real dilaton, the target space must have euclidean
signature $(G^{00}=+1)$ at the vicinity of fixed points with $c<25$,
and minkowskian signature $(G^{00}=-1)$ at the vicinity of fixed points with
$c>25$.
Ignoring small terms which are quadratic in first derivatives of the couplings,
eqs.(\ref{D2})--(\ref{D4})
  can be represented in the form
\begin{eqnarray}
  \ddot{\vec\lambda}+{Q}\dot{\vec\lambda} &=&
      \left\{ \begin{array}{ll} -\vec\beta & \mbox{for $c>25$} \\
      +\vec\beta & \mbox{for $c<25$},\end{array}\right. \ \ \ \hbox{with}
  \ \ \ Q^2 = {1\over3}\vert c-25\vert,\label{A4}\\
  \hbox{where}\ \ \vec\lambda & \equiv &
   \{G_{ij}(x),B_{ij}(x) \}.
\end{eqnarray}
Here, $\vec\beta$ are the $\beta$-functions (Weyl anomaly coefficients) of the
sigma model with $N$-dimensio-nal target space.

Next, consider a vicinity of a fixed point with $c=25$. As follows from
(\ref{A2}),
$Q(t)$ can no longer be approximated by a non-zero constant. Let us determine
$Q(t)$ to the next order in couplings.
Let us first argue that $\dot\varphi(t)$ can still be approximated by $-Q(t)$:
 According to (\ref{C3}),(\ref{C4}) the $\beta$-function for the
$\vec x$--dependent modes $\tilde\varphi(x)$ of $\varphi(x)$ has a linear piece
with plane waves
 $\tilde\varphi_p (\sim  a_p \cos px )$
  as eigenvectors. From eq.(\ref{D4}), $a_p(t)$ has
 some exponential $t$ dependence.
Eq.(\ref{D1}), expanded in Fourier modes, relates the amplitudes
$\tilde\varphi_p$
to the amplitudes of perturbations of $B$ and $G$. If the latter are
of order $\epsilon$, $\tilde\varphi_p$ is of order $\epsilon^2$ so that
 $\dot{\tilde\varphi}^2$ is of order $\epsilon^4$ and can be neglected in
(\ref{A3}). Eq.  (\ref{A3}) then takes the form
\begin{eqnarray}
  \dot Q=- \fourth \dot{\vec\lambda}^2\ ,\label{A5}
\end{eqnarray}
where $\dot{\vec\lambda}^2$ is defined by averaging as in (\ref{E1})
\begin{eqnarray}
  \dot{\vec\lambda}^2 \equiv
  <G^{ik} G^{jl} (\dot G_{ij} \dot G_{kl } +
  \dot B_{ij} \dot B_{kl })>.
\label{E2}\end{eqnarray}
Combining (\ref{A5}) with eq.(\ref{A2}) yields:
\begin{eqnarray}
 Q^2 = -{1\over3}G^{00}[\bar c(\vec\lambda)-25] + \fourth \dot{\vec\lambda}^2\
,
\label{A6}\end{eqnarray}
where the constant $c$ is now replaced by the $c$--function $\bar c$.
Whether $G^{00}$  should be chosen positive or negative now depends on the
behavior of $\bar c(\vec\lambda)$ at the vicinity of the fixed point.
Eq.(\ref{A6}) implies that if  $\dot \lambda$ is of order $\epsilon $ then $Q$
is  also of order $\epsilon$.
Thus,
\[\vert Q\vert=\vert\dot\varphi_0\vert  \sim O(\epsilon)\
   \gg\ \vert\dot{\tilde\varphi}\vert\sim O(\epsilon^2)\]
and so $\dot\varphi$  in eqs.(\ref{D2}),(\ref{D3}) can be approximated by
$-Q(t)$.\footnote{The fact that $Q \ge \epsilon$ while
 $\dot Q$ is of order $\epsilon^2$ signals the behavior $Q\sim{t^{-1}}$ (see
also the example in sec. 5).}
This yields:
\begin{eqnarray} \ddot{\vec\lambda}+{Q}\dot{\vec\lambda} +O(\dot\lambda^2) =
\left\{ \begin{array}{ll} -\vec\beta & \mbox{for $G^{00}=-1$} \\
    +\vec\beta & \mbox{for $G^{00}=+1$} \end{array}\right.
\label{A1}\end{eqnarray}
The $O(\dot\lambda^2)$ terms are
the $\dot G^2,\dot B^2,\dot G\dot B$ terms in (\ref{D2})--(\ref{D3}). Note that
they can be removed by redefining the variables:
for example, the $\dot G^2$ terms are removed
by changing variables to $\lambda^i_j$ such that $\dot\lambda^i_j=G^{ik}\dot
G_{kj}$ (in the basis where the metric
is diagonal, $G_{kj}=\delta_{kj}\ g_j,$ $\lambda^i_j=\delta^i_j\ \log g_j$).

The equations (\ref{A1}),(\ref{A4}) have been derived to leading order in
$\alpha'$. However,
at  the vicinity of fixed points  where time derivatives of $\vec \lambda$ are
small,
they are actually good approximations, with $\vec \beta$ containing all orders
in $\alpha'$ (see also sec. 4).

{}\section*{\lb 3. First application: attractive and repulsive fixed
points}\scs{3}

Depending on their initial parameters, classical solutions of
string theory are either attracted to fixed points or they diverge
as $t\rightarrow\infty$. By attractive fixed points we  understand static
solutions
that attract non--static ones within some {\it finite} volume domain of
initial parameters (by a finite -- as opposed to infinitesimal -- volume
domain we mean that no initial parameters are fine-tuned).
Let us now use the equations of the last section to  argue that
 attractive fixed points exist and correspond to minima of
$\vert \bar c(\vec\lambda)-25\vert$, where the $c$--function $ \bar c $ was
defined in (\ref{AA2}).
We begin with fixed points with $c\neq25$.
As shown, if a solution asymptotically approaches a
fixed point with $c\neq25$,  near the fixed point it obeys eq.(\ref{A4}), i.e.,
\begin{eqnarray}
 \ddot{\vec\lambda}+{Q}\dot{\vec\lambda}=
  \left\{ \begin{array}{lr}
  -\vec\beta & \mbox{for\ $c>25$},\\
  +\vec\beta & \mbox{for\ $c<25$},
  \end{array}\right. \ \ \hbox{with}\ \ Q^2\approx {1\over3}\vert
c-25\vert=\hbox{const} ,
\label{A7}\end{eqnarray}
To identify the attractors,  let us
 compare (\ref{A7}) with the  standard renormalisation group flow equation
for the $N$-dimensional theory,
\begin{eqnarray}
 \dot{\vec\lambda} = \left\{ \begin{array}{ll}
 -\vec\beta & \mbox{towards\ the\ IR},\\
 +\vec\beta & \mbox{towards\ the\ UV}.
\end{array}\right.
\label{A8}\end{eqnarray}
The time  in (\ref{A7}) is different from RG `time' in (\ref{A8}), but the
$\vec\beta$ functions are the same.\footnote{The  $\beta$-functions in
eq.(\ref{A8}) are  assumed to be defined
with specific `diffeomorphism terms' so that they represent the Weyl anomaly
coefficients \cite{ttt,hul}.}
This similar structure of eqs.(\ref{A7}),(\ref{A8})
leads to the following simple remarks (see also \cite{suss,tsey,mukh}
for some of the points; note also the discussion in ref.
\cite{ellis}):\footnote{It should be kept in mind that
since the order of these two equations is different, there are different
sets of initial conditions.  Also,  recall that we are ignoring the  gauge
constraints  present in the string system of equations.}

\begin{enumerate}
\item{\it Fixed points:}
Both equations describe a  motion of a particle in $\vec\lambda$--space with
the $\beta$-function as a driving force. They obviously
have the same fixed points, namely conformal field theories ($\vec\beta=0$).
\item{\it Sign of $Q$:}
If $Q$ is chosen to be positive, the motion (\ref{A7}) is damped.
If $Q$ is chosen
to be negative, it is {\it anti--}damped.
An anti--damped motion is the time--reversal of a damped motion.
It has no stable fixed points --
any small perturbation near a fixed point will blow up (sometimes preceded by
growing oscillations). Reaching a fixed point
would require fine--tuning the initial parameters.
In the following we are interested in stable fixed points of (\ref{A7});
for this purpose it is sufficient to consider the sector of solutions
with positive $Q$.
\item{ \it Stability of fixed points:}
Since the motion with positive $Q$ is damped, it follows from the
sign in front of $\vec\beta$ in (\ref{A7}), that the stable
fixed points of (\ref{A7}) with $c>25$ are the IR--stable
fixed points of (\ref{A8}). This is clear since
the only difference between  the two equations
is that the motion (\ref{A8}) is infinitely damped while (\ref{A7})
is finitely damped. Thus solutions of (\ref{A7}) may oscillate around
stable fixed points but will eventually settle down there.
On the other hand, the stable fixed points of (\ref{A7}) with $c<25$ are
the UV--stable fixed points of (\ref{A8}). To  summarise,
the stable fixed points of (\ref{A7}) are the minima of $\vert\bar c-25\vert$,
$\bar c$ being the (generalised) $c$--function \cite{zam,tsy}.
In other words, they are conformal field theories that contain no
relevant operators for $c>25$ or no irrelevant operators for $c<25$
(see  a discussion below).
\item{\it Consequence of the $c$--theorem }\cite{zam}:
Since (\ref{A7}) differs from the equation  for RG trajectories (\ref{A8}) only
by finite vs. infinite damping, it is clear that
cosmological solutions of (\ref{A7}) also obey an analog of the $c$--theorem:
if a solution  of (\ref{A7}) interpolates  between  two unitary CFT's,
an unstable  one  with a central charge $c_1$ and a  stable one with a central
charge $c_2$ and if both $c_1,c_2>25$
then $c$ decreases in the sense that $c_2<c_1$. If
$c_1,c_2<25$, $c$ {\it in}creases instead, $c_2>c_1$.
\item{\it Moduli directions:}
If a CFT has a  modulus, then (\ref{A8}) does not allow a flow in this
direction. By contrast, (\ref{A7}) allows `rolling' along the moduli
directions, which has been exploited in some cosmological string
solutions (see e.g. \cite{Mull},\cite{vene}).
For such solutions, $\bar c$ stays constant in time. In this
paper, the focus is instead on cosmological solutions with time--dependent
$\bar c$. For  solutions with $Q>0$, any `rolling' along moduli directions will
come to  rest after a while due to dilaton damping,
at least to leading  order in $\dot\lambda$.
\item{\it Limit $\vert c\vert\rightarrow\infty$:}
For $c\rightarrow\pm\infty$, $Q\rightarrow\infty$. In the case  when
$c\rightarrow+\infty$, a solution of (\ref{A7})
becomes  identical to the flow  (\ref{A8}) towards the IR  region if $2t/Q$ is
identified
with `RG time'. In the case
$c\rightarrow-\infty$,   the solution of
(\ref{A7}) is identical to the  RG flow towards the  UV region.

\end{enumerate}

Let us now also consider fixed points with $c=25$. Then $Q(t) $ is small and
not obviously positive,
 so the issue of stability is more subtle.
Let us show that IR or UV stable fixed points of the  RG flow are still stable
fixed points of the string equations of motion.  According to Sect.2.2,
in the vicinity of fixed points with $c=25$:
\begin{eqnarray}
  \ddot{\vec\lambda}+{Q}\dot{\vec\lambda} + O(\dot\lambda^2) &=&
G^{00}\vec\beta\  ,
  \label{G1}\\
  Q^2(t) &=& -{1\over3}G^{00}[\bar c(\vec\lambda)-25] +\fourth
\dot{\vec\lambda}^2,
  \label{G2}\\   \dot Q(t) &=& - \fourth \dot{\vec\lambda}^2.\label{G3}
\end{eqnarray}
The last equation states that $Q$ decreases, either to $-\infty$
or until a fixed point is reached.
Suppose the fixed point with $c=25$ is IR--stable and such that $\bar
c(\vec\lambda)-25$
is positive definite, and zero only at the fixed point.\footnote{I.e., we
assume the absence of moduli directions for simplicity.}
To have a real $Q$, $G^{00}=-1$ must be chosen. At the vicinity of the fixed
point,
consider a damped motion with $Q>0$. From (\ref{G3}),
$Q(t)$ will decrease and become 0 at some point. Then (\ref{G2}) implies
that the fixed point has been reached: $\dot{\vec\lambda}=0, c=25$. Eqs.
(\ref{G1}) and (\ref{G3}) imply that the trajectory will stay at the fixed
point:
$\dot Q=0, \ddot{\vec\lambda}=0, \ddot Q=0,...$, etc.
So IR stable fixed points of the RG flow with $c=25$ are also
fixed points of the above equations. Conversely,
IR unstable saddle points of the flow are also unstable fixed points of
the string equations: since $c(\vec\lambda)-25$ is not positive definite then,
there are solutions with $Q=0,\dot{\vec\lambda}\neq0\rightarrow\dot Q<0$,
so $Q$ becomes negative and the flow anti--damped. Analogous arguments
can be applied for Euclidean signature and UV stable fixed points. For an
example of a solution approaching a fixed
point with $c=25$, see sec. 5.

In summary, the attractors of time-dependent  string  solutions  are CFT's
with $c\ge25$ and no relevant operators, or CFT's with $c\le25$
and no irrelevant operators. One can always find `minisuperspace' examples
for such theories, as the one in sec. 5.
Beyond minisuperspace approximations, CFT's with no {\it irrelevant} operators
do not exist: in general,
perturbations of the form $\cos \vec p\vec x$ with arbitrarily high $\vec p^2$
and therefore dimension can be turned on.

On the other hand, CFT's with no {\it relevant} operators exist,
provided we ignore the non--derivative operators, corresponding to  the
tachyon.
At the
 linear order in the couplings, `no relevant operators' means `no operators
with
dimension less than two'. Such operators would correspond to tachyons
 in the low--energy effective
string theory and should be absent in all candidates for (super)string vacua.
To next order, the quadratic piece of the $\beta$-function of a marginal
(dimension 2) operator  must also be  non-negative. That this is a nontrivial
requirement will be seen on  the example in sec. 5.

{}\section*{\lb 4. Second application: RG flow in the presence of $2d$
gravity}\scs{4}

The similarity of the string equations of motion (\ref{A7}) and the
RG flow equations (\ref{A8}) has an interesting interpretation: special
solutions of (\ref{A7}) can be interpreted as renormalisation group
trajectories {\it in the presence of $2d$ gravity} \cite{das,suss,pkov}. Let us
briefly review this connection
and apply the preceding  discussion to conclude that the presence
of gravity does not allow to flow from a fixed point with $c>25$ to a
fixed point with $c<25$.\footnote{As explained in the introduction, we will not
worry about the tachyon for $c>1$ and  ghosts for $c\ge25$.}

Consider a  CFT with $N$ fields $x^i$, central charge $c$
(we choose $c\neq25$)
and world--sheet action $I_{\hbox{cft}}(\vec x)$, perturbed by interactions
$\Phi_i(\vec x)$ with small coupling constants $\lambda^i$ and
 scaling dimensions $h_i$:
\begin{eqnarray}
 I^{(N)}= I_{\hbox{cft}}(\vec x) +\lambda^i\int d^2 \xi \ \Phi_i(\vec x).
\label{J1}\end{eqnarray}
Without gravity, the $\beta$-functions for $\lambda^i$ are (see, e.g,
\cite{car})
\begin{eqnarray}
 \beta^i=(h_i-2)\lambda^i+\pi c^i_{jk} \lambda^j\lambda^k+...
\label{J2}\end{eqnarray}
with  $c^i_{jk}$ being OPE coefficients.
The model  (\ref{J1}) coupled to $2d$ gravity must be generally covariant, i.e.
 should be  described (in the  conformal
gauge \cite{d,dk})  by a {\it conformally invariant}
theory with an additional field $t$ related to the
conformal factor of the world--sheet metric \cite{polch,bankslykk,tseyt,suss}.
In the context of perturbation theory in $\lambda^i$  one finds to quadratic
order:\footnote{Here  target space gauge invariance has been used to eliminate
other terms with derivatives of $t$.}
\begin{eqnarray}
  I=I_{\hbox{cft}}(\vec x)+{1\over{8\pi}}\int  d^2\xi\ {\sqrt{\hat g}}[
\pm(\partial t)^2-
  Q R^{(2)} t ]  +\int d^2\xi \ \lambda^i(t) \Phi_i(\vec x)\ ,
\label{J3}\end{eqnarray}
\begin{eqnarray}
   \lambda^i(t)&=&\lambda^i\ e^{\alpha_i t}
  +{\pi\over{Q\pm2\alpha_i}}c^i_{jk}\lambda^j\lambda^k\ t e^{\alpha_i t}
  +O(\lambda^3), \label{J3.5}\\
  Q&=& \sqrt { {\vert 25-c\vert\ov 3}}  \ ,\label{J4}\\
  \alpha_i^2+Q\alpha_i &=&
      \left\{ \begin{array}{ll} h_i-2 & \mbox{for $c\ge25$} \\
      2-h_i & \mbox{for $c\le1$}.\end{array}\right.
\label{J6} \end{eqnarray}
The kinetic term for $t$ in (\ref{J3})
has a plus sign for $c\le1$ and a minus sign for $c\ge25$ in order to have real
$Q$.  The O($\lambda^2$) terms
in (\ref{J3.5}), introduced in \cite{sch}, are needed to insure conformal
invariance at quadratic order in $\lambda$ if there are nontrivial OPE
coefficients (see also \cite{pkk,amb,tanii}). By construction, $\vec\lambda(t)$
is a solution of the
conformal invariance equations (i.e.  string equations of motion in $N+1$
dimensions)
derived using perturbation theory in powers of couplings $\lambda^i$
near an $N$-dimensional conformal point.
Indeed, it is easily seen from (\ref{J6}) that $\vec\lambda(t)$ in
(\ref{J3.5}) obeys (\ref{A7}) up to order $\lambda^2$.\footnote{This confirms
again that (\ref{A7}) is actually  exact  near fixed points since it is
reproduced both in the two complementary approaches --
loop (or `power of $\alpha'$') expansion  and perturbative (or  `power of
coupling')   expansion.}

It is conjectured that $\lambda^i(t)$ in eq.(\ref{J3})
can be interpreted as running coupling constants of the  theory (\ref{J1})
coupled
to gravity. More precisely, $t$ can be
related to `renormalization group time' $\tau$ via the cosmological constant
operator\footnote{Such identification was considered also in \cite{ellis}.}
which defines the area of the surface, e.g,
\begin{equation}
  A=\int d^2\xi\ {\sqrt g}=\int d^2\xi\ {\sqrt{\hat g}}\ e^{\alpha t}
  =e^{-2\tau}\ \ \Rightarrow\ \ t=-{2\over \alpha}\tau,
 \ \ \alpha = -{Q\over2} + \left\{ \begin{array}{ll}
 {1\over2}{\sqrt{Q^2+8}} &\mbox{for\ $c\ge25$,}\\
 {1\over2}{\sqrt{Q^2-8}} &\mbox{for\ $c\le1$.}
\end{array}\right.
\label{K1}\end{equation}
$\tau$ is defined such that $\tau\rightarrow\infty$ corresponds to the UV.
Without coupling to gravity,  the coupling constant flow is determined by
(\ref{J2}).
With gravity switched on, one finds from (\ref{J3.5}):
\begin{equation}
  {d\over{d\tau}}\lambda^i=(\tilde h_i-2)\lambda^i
   +\pi\tilde c^i_{jk}\lambda^j\lambda^k+...
\label{K2}\end{equation}
with modified coefficients of the $\beta$-functions
\begin{eqnarray}
  \tilde h_i-2&=&-{2\over \alpha}\alpha_i,\label{K3}\\
  \tilde c^i_{jk}&=&-{2\over\alpha}{1\over{(Q+2\alpha_i)}} c^i_{jk}.
\label{K4}\end{eqnarray}
There is evidence that (\ref{J3.5}) indeed correctly describes the flow in the
presence
of gravity: (i) the phase diagram of the sine--Gordon model
coupled to gravity, derived in this way \cite{sch}, agrees with the matrix
model results \cite{moo}, and  (ii) the quadratic parts \cite{sch} of the
$\beta$-functions
in (\ref{K2})  for the case $\alpha_i=0$ agree \cite{amb} with the
light--cone gauge computation \cite{pkk}.

Let us determine which ends of the $t$--line correspond to
the infrared and the ultraviolet regions. At fixed points with central charge
$c$,
(\ref{K1}) yields:
\[ \alpha\ \hbox{is}\ \left\{ \begin{array}{ll} >0 & \mbox{for\ $c\ge25$,}\\
   <0 & \mbox{for\ $c\le1$} \end{array}\right.\ \ \ \Rightarrow\ \ \
   \hbox{IR corresponds to}\ \left\{ \begin{array}{ll} t\rightarrow+\infty
   & \mbox{for\ $c\ge25$,}\\ t\rightarrow-\infty
   & \mbox{for\ $c\le1$.} \end{array}\right.\]
For $1<c<25$, the relation between $t$ and $\tau$ is complex (we shall
 not  discuss this case here).

Since we have seen in section 3 that the attractive fixed
points of
string solutions are at $t\rightarrow+\infty$
($t\rightarrow-\infty$ corresponds to anti--damping), one might conclude that
there are no IR stable fixed points with $c\le1$ in the presence of gravity.
This is, however, not the case.
 The point is that only special solutions of
string theory can be interpreted as  RG
trajectories in the presence of $2d$ gravity. In  the theory of  $2d$
gravity
coupled to matter, we are instructed
to impose a boundary condition on $\lambda^i(t)$ in (\ref{J3}):
$\alpha_i$ in the couplings $\lambda^i(t)$ must have the values $\alpha_i^+$ in
\[ \alpha_i^\pm = \left\{ \begin{array}{ll}
 -{Q\over2}\pm{\sqrt{{Q^2\over4}-(h_i-2)}} &\mbox{for\ $c\ge25$,}\\
 -{Q\over2}\pm{\sqrt{{Q^2\over4}+(h_i-2)}} & \mbox{for\ $c\le1$,}
 \end{array}\right.\]
i.e.  it  must not take the `wrong' Liouville dressing values $\alpha_i^-$,
 corresponding to  `non--existing' operators and
states of Liouville theory \cite{sei,pch,anders}. This boundary condition is
required,
in particular,  for an agreement with matrix model
results and with the flow without gravity in the limit
$\vert c\vert\rightarrow\infty$.\footnote{In the matrix model,
one can turn on negative powers of matrices
in the matrix model potential, which have been conjectured to correspond
to the operators with the `wrong' Liouville dressing (see, e.g., \cite{dan}).
It would be very interesting to see if these negative powers have
an interpretation in terms of random lattices. Then the
additional boundary condition could be dropped and there would be no
distinction
between the flow in the presence of gravity and  string theory solutions. }

What does it mean to drop the negative Liouville dressings?
Consider a fixed point with $c<1$ and an irrelevant direction, $h_i>2$;
 in that case  the flow {\it without} gravity converges to a fixed point in the
IR.
Then $\alpha^+_i>0,\  \alpha^-_i<0$, so that $e^{\alpha^+_it}$ goes to zero  in
the IR
($t\rightarrow-\infty$), while $e^{\alpha^-_it}$ diverges in the IR.
For string solutions, both  choices are allowed.
The statement that the motion is anti--damped and that the fixed point is
therefore unstable at $t\rightarrow\infty$
is precisely the statement that the general linear
combination of both solutions diverges at $t\rightarrow\infty$.
But for the RG flow in the presence of gravity  the solutions that contain
$\alpha^-_i$ are not allowed.
Under this restriction, the flow does have IR stable fixed points with $c\le1$:
as without gravity, they are  minima of $\bar c$.

One may then wonder whether there are any qualitative
changes in the RG flow due to coupling to  $2d$ gravity, apart
from the  quantitative modification of the $\beta$- function coefficients. Let
us conclude by pointing out two qualitative effects.
 First, being second order in
derivatives, the flow  in the presence of gravity can oscillate around fixed
points before settling down at them
(cf. (\ref{A8}) and (\ref{A7})).
As a consequence, the  $\bar c$--function
 does not strictly decrease along RG trajectories.\footnote{Note that the
$c$-theorem in its original formulation \cite{zam}
 does  not  apply to   sigma models  and  in the
presence of (non-unitary) dilaton (or ghost) interaction terms,
see  \cite{tsy,CP,os,cth}.}
Using
(\ref{A7}) and the ordinary $c$--theorem ($\dot{\bar c}\le0$ towards the IR)
and writing $\dot{\bar c}\sim\dot\lambda^i{\partial \bar c /
\partial\lambda^i}$,
it is easy to derive the modification of the $c$--theorem by gravity
at the vicinity of fixed points with central charge $c$:
\[ \ddot{\bar c} + Q\dot{\bar c}\
    \left\{ \begin{array}{ll} \le0 & \mbox{for $c>25$} \\
      \ge0 & \mbox{for $c\le1$},\end{array}\right. \ \ \ \hbox{with}
   \ \ \ Q^2={1\over3}\vert c-25\vert.\]
Here, the dot represents derivative with respect to $t$, not $\tau$.

A second  qualitative effect of gravity concerns flows that start
from $c>25$ but  do not reach   a fixed point with $c\ge25$. They obey
eq.(\ref{A2}), which is valid everywhere
(i.e., not only near  fixed points) up to order $\alpha'$:
\[  \dot Q = {{c(\vec\lambda)-25}\over3} - Q^2.\]
Note that if  $Q$ is real, it stays real at all times. Since
$Q^2$ would have to be negative at fixed points with $c<25$,
$\vec\lambda(t)$ cannot converge to a CFT with $c<25$.
It {\it is} possible for $\bar c(\vec\lambda)$ to become smaller than 25, but
then $\dot{\vec\lambda}$ must be large enough to make $Q^2\ge0$ in (\ref{A6}).
We conclude that in the models with dynamical $2d$ gravity there is a
`$c=25$ barrier' for the flow: a flow that starts with $c>25$
is either attracted to a fixed point with $c\ge25$, or it does not converge to
a fixed point at all (see also the example in
the next section).
This result, derived here to the leading order in  $\alpha'$, may  be true to
all orders.

{}\section*{\lb 5. An example: Group space }\scs{5}

The purpose of this section is to illustrate some of the preceding  discussion
on
 a `minisuperspace' example, considered previously in \cite{Anto}, with
solutions which   interpolate
between a flat space and  a  group  space (the Wess--Zumino--Witten model).
Consider the sigma model with a WZW term \cite{wzw}. The
$N$--dimensional target space is the group space  $G$  with a
fixed metric $\hat G_{ij}$  and curvature
\[  {\hat R}^{(N)}_{ij}={1\over4}f_{imn}f_j^{\ mn}
  ={1\over4}c_G\ \hat G_{ij},\]
where $f_{ijk}$ are the structure constants and $c_G$ is the value of the
quadratic
Casimir operator in the adjoint representation.
The WZW term   can be represented by
the  antisymmetric tensor field $\hat B_{ij}$ with  the field strength
\[  \hat H_{ijk}=k\ f_{ijk} .\]
For the time dependence of the background we make the ansatz
\begin{eqnarray}
  G_{ij}(\vec x,t)&=& {1\ov g^2 (t)} {\hat G}_{ij}=
e^{2\lambda(t)}{\hat G}_{ij}
  (\vec x),\\
  \phi(\vec x,t)&=&\phi(t),\\
  H_{ijk}&=&\hat H_{ijk} = k \ f_{ijk}.
\end{eqnarray}
This  ansatz is consistent as a consequence of the group symmetry;
one can also show that, as expected,
 there are no solutions with time-dependent $k$.
$\lambda$ and $\phi$, or  $\lambda$ and
$\varphi=2\phi-N\lambda$ are the coupling constants of the  model.
For the minkowskian sector of solutions with $G^{00}=-1$
the equations for $\lambda(t)$ and $Q\equiv -\dot\varphi(t)$ are
(to order $\alpha'$) \cite{tsey}
\begin{eqnarray}
  \ddot\lambda+{Q}\dot\lambda &=&
  -{1\over{6N}}\ {\del \bar c(\lambda)\over \del \lambda }\  ,\label{H0}\\
  Q^2 &=& N\dot\lambda^2+{1\over3}[\bar c(\lambda)-25]\ , \\
   \dot Q&=&-N\dot\lambda^2\  .
\label{H1}
\end{eqnarray}
Here  the $\bar c$--function  (\ref{AA2}),(\ref{C4})  is
\begin{eqnarray}
 \bar c(\lambda)= C^{(N)}= \ N-3(R^{(N)}-{1\over12}H^2)\ = \
  N+{3\over4}Nc_G (-e^{-2\lambda} +{1\over3}k^2e^{-6\lambda}).
\label{HH}
\end{eqnarray}
For comparison, the standard
RG flow towards the IR region in the $N$-dimensional model
 is  determined by
\be\label{QQ1}
 \dot \lambda =\beta(\lambda) = -{1\over{6N}}\  {\del\bar c(\lambda)\over
\del\lambda }= - \fourth c_G e^{-2\lambda}(1- k^2e^{-4\lambda}) \ .
\ee
Near fixed points (\ref{H0})--(\ref{H1}) agree with
(\ref{A5}),(\ref{A6}),(\ref{A1})
(up to a normalisation of $\lambda$; $\dot\lambda^2$ terms are absent in this
case).
$\bar c(\lambda)$ approaches $N$ from below for
$\lambda\rightarrow\infty$ ($\lambda =\infty, \ \bar c= N$ is a trivial fixed
point)
   and has the  minimum (corresponding to the well-known  zero of $\beta
(\lambda)$\cite{wzw})  at
\[  e^{2\lambda} = |k|\  ,  \ \ \ \ \bar c_{min}=N-{c_GN\over 2|k|}
 +O({1\over k^2}) =  {{N|k|}\over{|k|+\ha c_G}}. \]
We are restricted to large enough $\lambda$ so that higher order
corrections in $\alpha'$ can be neglected.

The above equations illustrate the following points of sec. 3 (points 1,2 and
4) and of sec. 4 (point 3):
\begin{enumerate}
\item
Eqs (\ref{H0})--(\ref{H1}) with $Q>0$
describe a damped motion of a particle in the potential $\bar c(\lambda)$.
The model can be viewed as interpolating between different internal spaces,
namely, a  flat space (the asymptotically free limit $\lambda\rightarrow\infty$
of the sigma model) and a curved space (the WZW model).
If $\bar c_{min}>25$, $\lambda$ settles down -- after possible oscillations --
at the WZW  fixed point  (`compactification'), while for $\bar c_{min}< N <25$
(corresponding to the Euclidean sector of solutions) the model
approaches (at large Euclidean time) the flat space (`decompactification' in
Euclidean time).
\item
If $c_{min}=25$, the solutions are still attracted to the WZW model. In its
vicinity (where the potential $\bar c$ is proportional
to $\lambda^2$ where $\lambda$ has been shifted such that it is zero at the WZW
point), $\vec\lambda(t)$ performs
oscillations with an amplitude $\Lambda(t)$ and the energy $Q^2(t)$ which
satisfies
\begin{eqnarray}
  Q^2 = E_{kin}+E_{pot},\ \ \ \  {d\over{dt}}Q^2= -2N Q \dot{\lambda}^2,\\
  E_{kin} = N \dot{\lambda}^2 , \ \ \
  E_{pot} = {1\over3}[\bar c(\lambda)-25] .
 \end{eqnarray}
Due to the damping, $\Lambda$ and $Q$ can be seen to slowly decrease with time
: $ \  Q,\Lambda\sim{ t^{-1} }. $
\item
If $N>25$ but $\bar c_{min}<25$, there is no stable fixed point. Indeed,
starting from the flat space model with $\bar c=N$,
eq.(\ref{H1}) implies that $Q$ decreases until a new fixed
point is reached. But the WZW fixed point with $\bar c_{min}<25$
cannot be reached since it would correspond to imaginary $Q$,
and there is no other fixed point with $c\ge25$. So $Q$ keeps decreasing,
becomes negative, and the  solution  blows up due to anti--damping\footnote{A
similar remark applies to solutions that start from the WZW fixed point with
increasing $\bar c$.}
(of course, at some point terms of higher orders  in $\alpha'$ must be taken
into
account). Applied to the RG flow in the presence of $2d$ gravity,
this illustrates that starting from a fixed point with $c>25$ the
flow  cannot reach  another fixed point with $c<25$ even  though this  was
possible
in the absence of coupling to  gravity.  Instead, the flow diverges.
\item
Near the WZW fixed point, the $\lambda$-perturbation has  dimension $>$ 2, so
for $c_{min}>25$ the WZW point  is an attractor
 within this `minisuperspace'.
The other fixed point, corresponding to the free model,  is unstable although
it has no operators with
dimension less than two (the $\lambda$-perturbation has dimension two
at the free end). The reason is  a negative quadratic  $\beta$-function or OPE
coefficient.
The requirement of not having negative quadratic beta functions
for operators with dimension two is thus a nontrivial restriction
on attractors.

\end{enumerate}

{}\section*{\lb 6. Concluding  remarks}\scs{6}

Viewing cosmological solutions of  string theory as trajectories
in the space of $2d$  field theories, we have  seen that
while the trajectories diverge for some domains of initial values, there
are domains within which the trajectories
are attracted to the minima of $\vert \bar c-25\vert$.
In particular, minima with $\bar c=25$ are  also attractors.
As mentioned in the introduction, one hope is to
assign, already at the classical level, different probabilities
to different string vacua, proportional to the sizes of the
corresponding domains of attraction.

The present discussion cannot explain why the universe is not stuck, e.g.,
in a minimum of $\bar c$ with $\bar c>25$. But our analysis  was  restricted to
classical string theory. One
may speculate that in the quantum theory the universe can tunnel between
various minima of $\vert \bar c-25\vert$ until it reaches a ground state with
$\bar c=25$.

Quantum effects in string theory  are  described by sums over world sheets of
different topologies.
In view of the conjectured interpretation of (some of the) cosmological string
solutions
as renormalisation group trajectories in the presence of  dynamical $2d$
gravity,
one may wonder whether the only IR--stable fixed points of two--dimensional
field theory on a surface with fluctuating geometry {\it and} topology are {\it
critical} ($c=25$)  string vacua.

\vskip1mm\noindent{\bf Acknowledgements:}

\noindent
The work of  C.S.  is supported by Schweizerischer Nationalfonds.
A.A.T. would like to acknowledge the support of  PPARC.

\baselineskip=12pt

{}

\end{document}